\documentclass[twocolumn,aps,prl,superscriptaddress,multicol,amsmath,amssymb,showpacs,10pt]{revtex4-2}
\usepackage[utf8]{inputenc}         
\usepackage{microtype}              
\usepackage{array}                  
\usepackage{booktabs}
\usepackage{amsmath,amssymb}
\usepackage{times}
\usepackage{siunitx}
\usepackage{graphicx}

\usepackage[dvipsnames]{xcolor}
\usepackage{sidecap}

\usepackage[colorlinks,linkcolor=blue,citecolor=blue,anchorcolor=blue]{hyperref}
\usepackage{cleveref}
\usepackage{nameref,zref-xr}
\zxrsetup{toltxlabel}

\usepackage{textcomp}
\begin{document}

\title{Orientation-driven route to an intrinsic insulating ferromagnetic state in manganite superlattices}

\author{Priyanka Aggarwal}
\thanks{These authors contributed equally to this work.}
\affiliation{Department of Physics and Astronomy, Uppsala University, Box 516, SE-75120, Uppsala, Sweden}
\affiliation{Asia Pacific Center for Theoretical Physics, Pohang, Gyeonbuk 790-784, Republic of Korea}

\author{Kirill B.\ Agapev}
\thanks{These authors contributed equally to this work.}
\affiliation{Physical Science and Engineering Division, King Abdullah University of Science and Technology (KAUST), Thuwal 23955-6900, Kingdom of Saudi Arabia}

\author{Sagar Sarkar}
\affiliation{Department of Physics and Astronomy, Uppsala University, Box 516, SE-75120, Uppsala, Sweden}

\author{Biplab Sanyal}
\affiliation{Department of Physics and Astronomy, Uppsala University, Box 516, SE-75120, Uppsala, Sweden}

\author{Igor Di Marco}
\email[I. Di Marco: ]{igor.dimarco@physics.uu.se}
\affiliation{Institute of Physics, Nicolaus Copernicus University, Grudziadzka 5, 87-100, Toru\'n, Poland}
\affiliation{Department of Physics and Astronomy, Uppsala University, Box 516, SE-75120, Uppsala, Sweden}

\author{Fabrizio Cossu}
\email[F. Cossu: ]{fabrizio.cossu@york.ac.uk}
\affiliation{School of Physics, Engineering and Technology, University of York, Heslington, York YO10 5DD, United Kingdom}
\affiliation{Department of Physics and Institute of Quantum Convergence and Technology, Kangwon National University, Chuncheon, 24341, Republic of Korea}
\affiliation{Department of Physics, School of Natural and Computing Sciences, University of Aberdeen, Aberdeen, AB24 3UE, UK}

\date{\today}

\begin{abstract}
	Increasing precision in the growth of superlattices sparks hope in applications that may arise from engineering layered structures. Heterostructuring and functionalization of magnetic oxides have been very popular due to their versatility and readiness for integration in modern electronics. In this study, we provide yet another example of this phenomenology by predicting that an insulating ferromagnetic state can be realized in superlattices of LaMnO$_3$ and SrTiO$_3$ oriented along the (111) direction. In strike contrast with respect to other orientations, these properties are not of extrinsic origin but arise from the interplay of structural order, strain and quantum confinement. The bandgap is shown to be either direct and indirect, depending on the precise composition, which can be explained in terms of the geometrical properties of (111)-oriented bilayers of LaMnO$_3$.
    The electronic structure shows narrow bands indicating localized $e_g$ states for all the investigated superlattices. These features and the analysis of the inter-atomic magnetic coupling suggest that the investigated superlattices behave as a Kugel-Khomskii material, at least for the explored compositions. Our results provide not only a new route to an insulating ferromagnet, but also novel insight into the intricate interplay between lattice symmetry, Hubbard physics and Hund's coupling to be exploited in next-generation spintronic applications.

\end{abstract}


\maketitle 

 The orbital physics of $d$ states in transition‐metal oxides has long been recognised as a key factor in driving the rich phenomenology of these correlated electron systems.
 The intertwining between lattice, charge, orbital and spin degrees of freedom manifests in different ways.
 While partially filled $t_{2g}$ orbitals give rise to exciting phenomena such as itinerant magnetism and frustrated magnetism, $e_{g}$ orbitals govern processes such as double‐exchange and superexchange, and strongly influence structural distortions~\cite{varignon-PRR.1.033131} and quantum confinement~\cite{varignon-ncomm2019} effects.
 The symmetry and spatial distribution of these intertwined degrees of freedom can be finely tuned through epitaxial growth, which results in electronic and magnetic states different from the bulk~\cite{hellman2017rmp}. 
 For instance, a very interesting phenomenology can be obtained in heterostructures of LaMnO$_3$ and SrTiO$_3$. 
 LaMnO$_3$ is a paradigmatic example of a charge-transfer insulator, where the singly occupied $e_g$ electron favors Jahn-Teller distortions and thus an A-type antiferromagnetic (AFM) order~\cite{schmitt-PRB.101.214304}. SrTiO$_3$ is instead a non-magnetic band insulator that is commonly used for substrates in heterostructures of perovskite oxides~\cite{Pai_2018}. When thin films of LaMnO$_3$ are grown on SrTiO$_3$ along the (001) direction, a ferromagnetic (FM) insulator with a relatively high T$_c$ of 140~K can be obtained~\cite{Choi_2009,wang_xr-Science2015,anahory-ncomm2016}. A similar state can also be obtained in (001)-oriented superlattices of LaMnO$_3$ and SrTiO$_3$~\cite{choi-PRB.83.195113,zhai13jap113_173913,zhai-ncomm2014,chen2020appmatint}. FM insulators are relatively rare, yet they are key building blocks in spintronic architectures, e.g. as spin-filter barriers and spin-polarizing elements, and are relevant for nonvolatile memory concepts~\cite{bader_spintronics}. At a fundamental level, they are also intriguing for quantum computing and topological effects~\cite{liu_review_mag_top}.
 Unfortunately, candidate materials usually have a low Curie temperature and often exhibit air and chemical sensitivity, pronounced thickness dependence (for layered systems), and structural fragility~\cite{ferroins1,ferroins2,ferroins3}. 
 %
Thus, the perspective of realizing novel FM insulating states in LaMnO$_3$ thin films and heterostructures was received with great interest from the scientific community, excited also by the possibility of an easier integration of perovskites into modern oxide-based technology~\cite{PhysRevLett.134.016702}. 
Nevertheless, a scientific consensus has slowly emerged that the FM insulating state observed in the aforementioned experiments is not explained by an intrinsic reconstruction of an ideal, stoichiometric LaMnO$_3$ film alone~\cite{lee_jh-PRB.88.174426}. Instead, it likely reflects an interplay between interfacial driving forces and defect-mediated compensation. The polar discontinuity at the polar/nonpolar interface provides an intrinsic electrostatic driving force for charge redistribution~\cite{wang_xr-Science2015,chen_zh-PRL.119.156801,PhysRevB96235112,li2019advmat}, which in real samples may be accommodated through thickness-dependent oxygen nonstoichiometry and cation intermixing~\cite{niu20181800055,li2019advmat,chem2020prm4,niu2021advelemat,folkers2024prm,stramaglia2024aipadv}, thereby promoting ferromagnetism and dead layers. The insulating character has been interpreted in terms of electronic/magnetic inhomogeneity, for example via the formation of isolated magnetic nanoislands~\cite{anahory-ncomm2016,PhysRevB96235112}.
Thus the question of creating a perovskite heterostructure that is uniformly ferromagnetic yet insulating in the intrinsic, defect-minimized limit remains open.

\begin{figure}[tb]
\centering
	\includegraphics[trim = 0 0 0 0,width=1.0\linewidth]{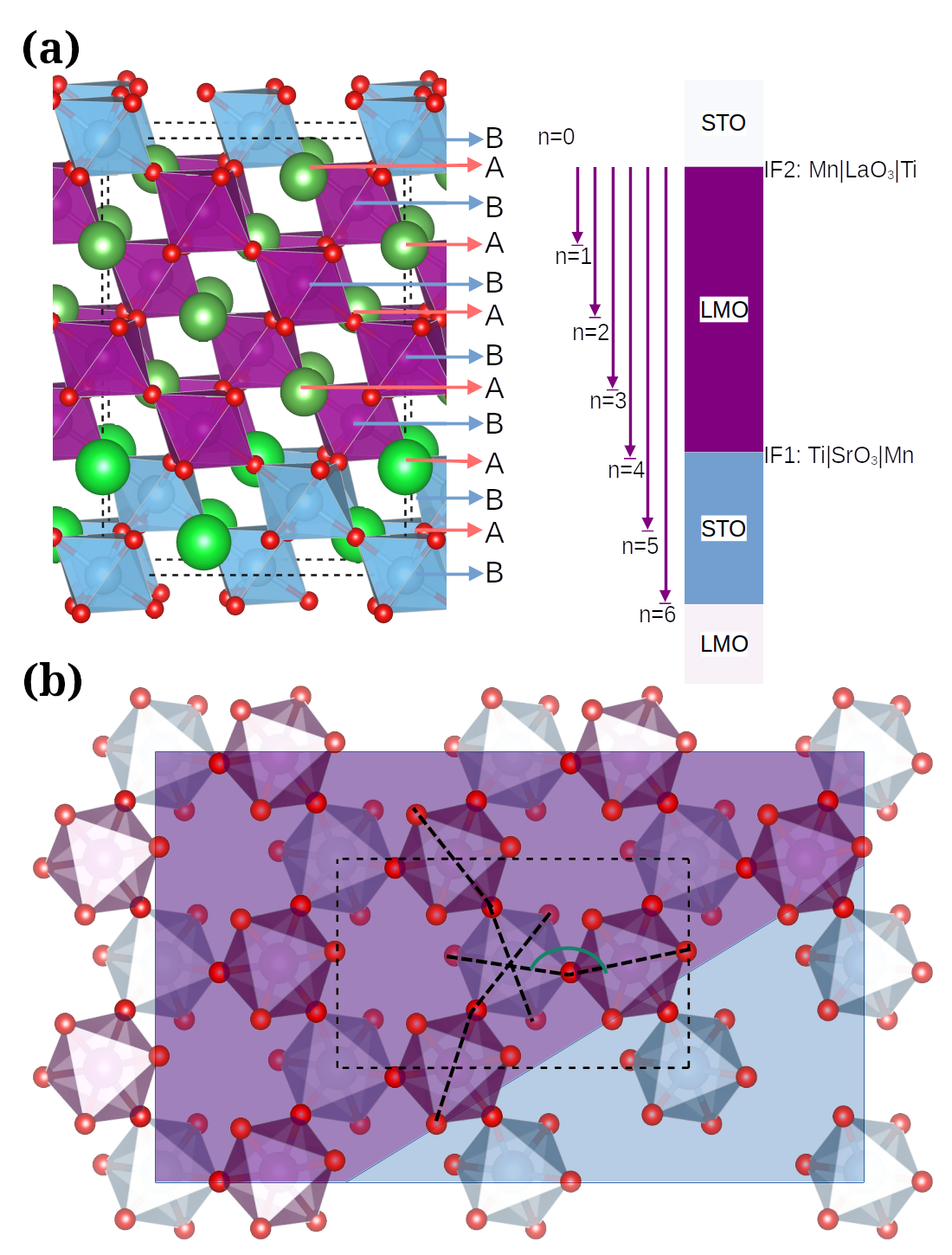}
	\caption{(a) stacking order of (111)-oriented superlattices of ABO$_3$ perovskites (green spheres A=Sr/La; purple spheres B=Ti/Mn; red spheres O), accompanied by a sketch illustrating the considered thickness of the component regions LaMnO$_3$ and SrTiO$_3$, as well as the hetero-interfaces between them (IF1, IF2); (b) schematic illustration of two trigonal (111) layers composing a buckled honeycomb lattice for the joint bilayer.}
\label{fig:sketch}
\end{figure}

In the last decade, significant effort has been devoted to investigate oxide heterostructures that are oriented along the (111) direction, in particular for nickelates~\cite{chakhalian-RMP.86.1189,chakhalian-APLMat2020,okamoto_prm_2025}. Manganites as well have been investigated, following the prediction that a LaMnO$_3$ bilayer grown on a (111) substrate should exhibit non-trivial topological properties~\cite{xiao-ncomm2011,weng-PRB.92.195114,tahini-PRB.93.035117,okamoto_jpsj_2018}. This is a consequence of the geometry of (111) perovskite bilayers, which \textit{de facto} correspond to a buckled honeycomb lattice of Mn atoms, as illustrated in Figure~\ref{fig:sketch}(b). In these settings, quantum confinement and strain are expected to modify structural, electronic and magnetic properties profoundly~\cite{tahini-PRB.93.035117}. Recent experimental works have also demonstrated that these bilayers are possible to synthesize~\cite{jansen2024prm}, confirming the feasibility of (111)-oriented heterostructures of manganites~\cite{zhou_y-AMI2019,xu2023}.
The properties of (111) superlattices are even more intriguing,
as different band topologies may be tuned by allowing for combinations of an even/odd number of layers in the component regions.
Therefore, these systems represent a versatile platform to explore novel phenomena associated to correlated topological states that would not be accessible for the (001) orientation~\cite{doennig2016}.

 In this letter, we predict that an intrinsic FM insulating state can be obtained in (111)-oriented superlattices of LaMnO$_3$ and SrTiO$_3$ for various periodicities. This state is not driven by interfacial effects, but arises from structural order, strain and quantum confinement typical of a Kugel-Khomskii material. The direct/indirect nature of the bandgap is shown to have a non-trivial dependence on the thickness of the component regions, which is explained in terms of the particular geometrical properties of (111)-oriented bilayers with respect to stand-alone monolayers. The electronic structure is characterized by narrow bands that indicate localized $e_g$ states for all the studied compositions. Our results  provide insight into the intricate interplay between lattice symmetry, electron correlations, and Hund's coupling, while also suggesting new avenues for exploiting these phenomena in next–generation spintronic applications.

\textit{Ab-initio} electronic structure calculations of (111)-oriented superlattices of chemical formula (SrTiO$_3$)$_{6-n}\vert$(LaMnO$_3$)$_{n}$, with $n=1-5$, are performed by means of the Vienna {\itshape{ab-initio}}
 simulation package (VASP)~\cite{kresse-PRB.54.11169,kresse-PRB.59.1758}. The overall thickness of 6 layers  represents the minimal thickness along the (111) direction that is compatible with the space groups to be considered. 
 The in-plane lattice constant with respect to the (111) direction is fixed to model epitaxial growth on a (111) substrate of cubic SrTiO$_3$~\cite{ARYA2023111917}, using the theoretical lattice constant of \SI{3.925}{\angstrom}.
The exchange-correlation functional is treated in the generalized gradient approximation (GGA)~\cite{perdew-PRL.77.3865,perdew-PRLerratum.78.1396}. The Mn-$3d$ and Ti-$3d$ states are treated within the rotationally invariant DFT+$U$ approach by Liechtenstein \textit{et al.}~\cite{liechtenstein-PRB.52.r5467,kotliar2006}. Coulomb interaction parameters are set to $U =$ \SI{4.0}{\electronvolt} and $J =$ \SI{1.0}{\electronvolt} for Mn and to  $U =$ \SI{5.0}{\electronvolt} and $J =$ \SI{0.5}{\electronvolt} for Ti, which previous literature demonstrated to reproduce the bulk and heterostructure properties to a good accuracy~\cite{nanda-PRB.81.224408,nanda-PRL.101.127201,chen_zh-PRL.119.156801,myself-EPL2017,myself-npjCM2022,jansen2024prm,liu2025prb,mellan-PRB.92.085151,Banerjee2019prb,myself-npjCM2022,PhysRevLett.116.157203,PhysRevB.96.161409,PhysRevLett.119.256403}. Further details on the calculations and the post-processing software~\cite{bader-AccChemRes1985,VESTA-JACr2011,VASPKIT-1908.08269} are reported in the Supplementary Material (SM)~\cite{supplemental}.

 The superlattices are constructed from the $Pnma$ and $R\bar{3}c$ bulk structures, featuring $a^{-}a^{-}c^{+}$ and $a^{-}a^{-}a^{-}$ tilting systems (in Glazer's notation \cite{glazer-AC:B1972}), respectively. Full structural relaxation based on energy and stress tensor minimisation~\cite{myself-npjCM2022} is then performed for the four most favorable magnetic phases, namely FM and AFM of types A, C and G, whose depiction along the (111) direction is shown in Figure~S1 of SM~\cite{supplemental}. For simplicity, we will refer to the three AFM orders as A-AFM, C-AFM and G-AFM, while we will use the term ``spin'' to indicate the local spin magnetic moments. 
\begin{table}[b]
   \centering
    \caption{\label{toten_strc-magn:tab} Relative energy of competing magnetic states in the most favorable tilting patterns for (111)-oriented superlattices of chemical formula (SrTiO$_3$)$_{6-n}\vert$(LaMnO$_3$)$_{n}$, where $n =1-5$.
    Values are given in meV per formula unit and with respect to the GS for any given $n$.
    No magnetic solution could be found for $n=1$ with the $a^{-}a^{-}c^{+}$ tilting pattern.}
    \begin{tabular*}{0.8\linewidth}{>{\raggedright\arraybackslash}p{1.2cm} >{\centering\arraybackslash}p{1.0cm} >{\centering\arraybackslash}p{1.0cm} >{\centering\arraybackslash}p{1.0cm} >{\centering\arraybackslash}p{1.0cm} >{\centering\arraybackslash}p{1.0cm}}
\toprule
       & 5-1 & 4-2 & 3-3 & 2-4 & 1-5 \\
\midrule
   \multicolumn{6}{c}{$a^-a^-a^-$ ($R\bar{3}c$)} \\
\hline
 A-AFM &    GS     & 6.1  &  15.5 &  20.8  &  113.5  \\
 C-AFM &    GS     & 16.7 &  15.2 &  25.5  &  122.4  \\
    FM &    5.6     & 4.0  &  4.6  &  7.1   &  18.3  \\
\midrule
       \multicolumn{6}{c}{$a^-a^-c^+$ ($Pnma$)} \\
\hline
 A-AFM &    --     & 4.3  &   5.4  &  7.7  &   96.2  \\
 C-AFM &    --     & 8.7  &   25.6 &  21.9 &   108.1  \\
    FM &    --     &  GS    &    GS    &    GS   &    GS    \\
\bottomrule
    \end{tabular*}
\end{table}
\begin{figure*}[tb]
\centering
	\includegraphics[trim = 0 0 0 0,width=0.95\linewidth]{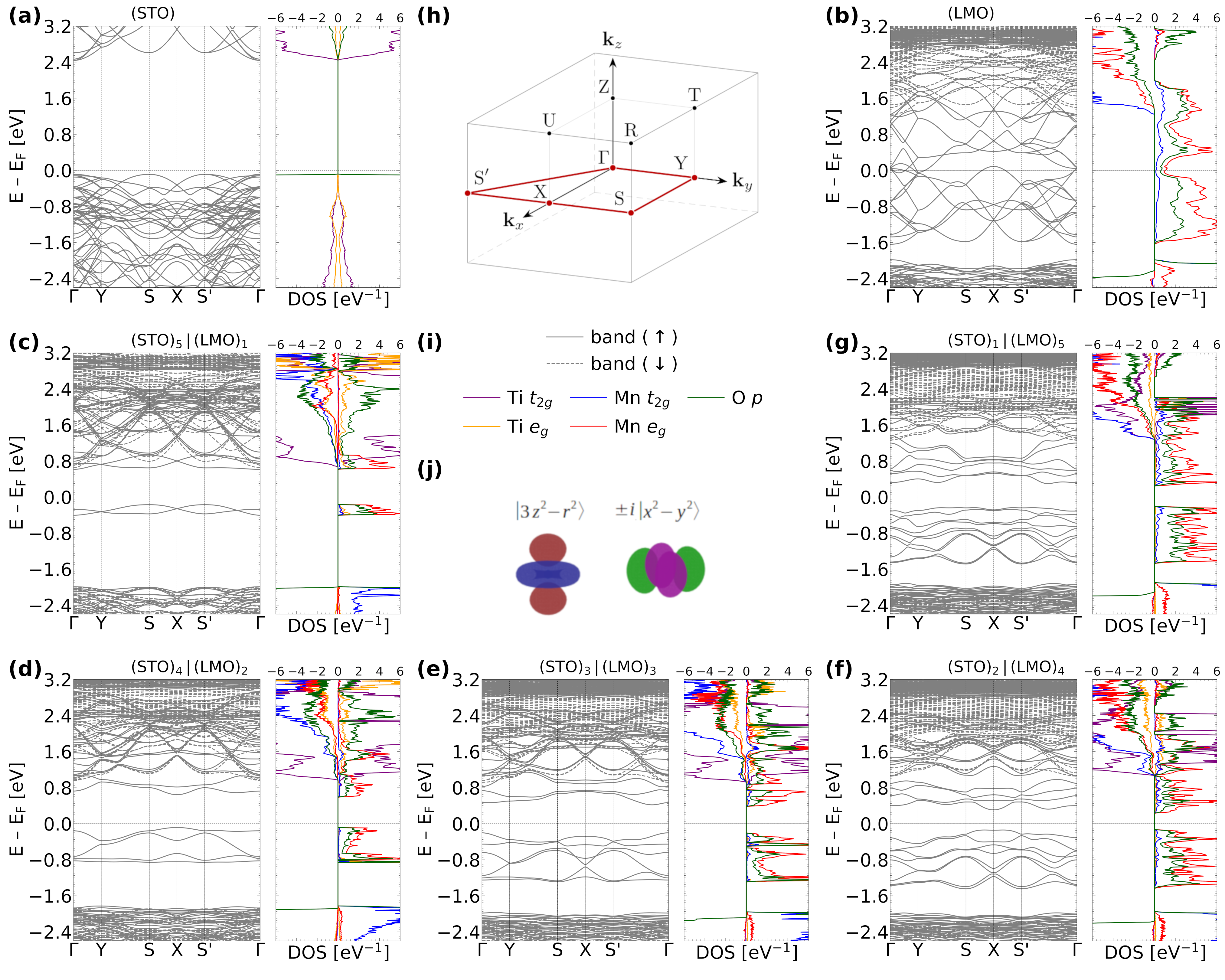}\\
	\caption{\label{fig:bandos} Panels (a-g): band structure and projected density of states (PDOS) for the investigated superlattices in their FM GS. Panel (h): high-symmetry paths in the Brillouin zone. Panel (i): legend. Panel (j): $e_g$-orbital linear combinations defining the orbital ordering along the (111) direction~\cite{chakhalian-APLMat2020}, which select ligand-mediated superexchange pathways in the Kugel–Khomskii picture~\cite{kliment-SPU1982,feiner1999prb,Khomskii_2022}.
    }
\end{figure*}
 The energies of the FM, A-AFM and C-AFM phases are reported in Table~\ref{toten_strc-magn:tab}. The G-AFM order is found to be significantly higher in energy and therefore is not even shown in the Table. The ground state (GS) is FM with $a^-a^-c^+$ tilting pattern for all compositions but $n=1$. For the latter, a transition to a AFM state with $a^-a^-a^-$ tilting pattern is observed. We notice here that for $n=1$ there are not enough layers to differentiate between different types of AFM order and thus this is a simple in-plane AFM system (see also Section~II of SM~\cite{supplemental}).
 The visual inspection of the structure demonstrates that the $a^-a^-c^+$ tilting pattern obtained for $n=2$ and $n=3$ is closer to the $a^-a^-c^0$ configuration, characterized by the absence of octahedral rotations and nearly straight Mn(Ti)–O–Mn(Ti) bond angles along the local $z$-axis. This intermediate order is a precursor of the composition-driven structural transition to the $a^-a^-a^-$ tilting pattern.
 The perfect correspondence between structural and magnetic transition observed in Table~\ref{toten_strc-magn:tab} is not a physically relevant result, since the precise threshold of both transitions depend on the parameters used, namely the value of the in-plane lattice constant describing the epitaxial growth and the Coulomb interaction parameters. As shown in Section~VI of SM~\cite{supplemental}, for a slightly different set of parameters, the structural transition precedes the magnetic transition, while the overall trend of magnetic and structural stability remains the same; for sufficiently thick regions of LaMnO$_3$, a robust FM state is obtained. Finally, notice that for $n=5$, the A-AFM and C-AFM solutions are substantially higher in energy. This behavior can be attributed to the presence of only a single SrTiO$_3$ layer separating adjacent LaMnO$_3$ layers, which enables a residual magnetic interaction across the spacer, favoring the FM alignment strongly.
\begin{figure*}[t]
\centering
	\includegraphics[trim = 0 0 0 0,width=1.0\linewidth]{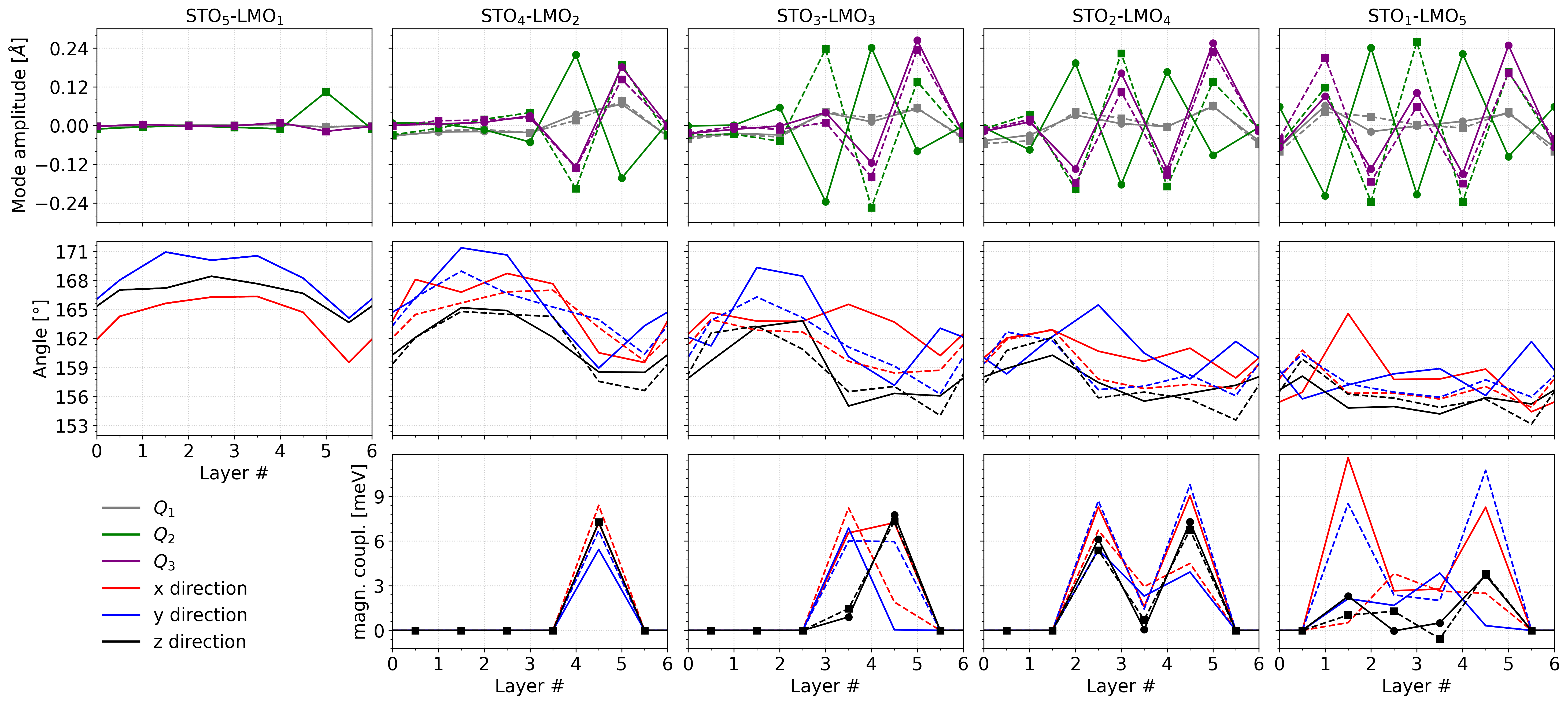}
	\caption{Layer-resolved van Vleck distortions~\cite{supplemental} (top panels), Mn-O-Mn, Mn-O-Ti and Ti-O-Ti angles (middle panels) and nearest-neighbor magnetic interactions (bottom panels) of the five investigated superlattices in the FM GS. Continuous and dashed lines refer to two distinct sublattices arising for the $a^-a^-c^+$ tilting pattern~\cite{myself-PRB.109.045435}. A positive sign of the magnetic coupling corresponds to a parallel FM alignment.}
\label{fig:angmgn}
\end{figure*}

The electronic properties of the superlattices in their GS are illustrated in Figure~\ref{fig:bandos}, alongside the corresponding data for pure SrTiO$_3$ and LaMnO$_3$ in the superlattice geometry. For sake of comparison, we are going to discuss the FM state also for $n=1$, albeit the real GS is AFM. To avoid ambiguities we will thus refer to the FM GS from this point.
All mixed superlattices are insulating, pointing to electronic state localization, which has been previously reported in analogous (111)-oriented systems~\cite{dong-PRB.87.195116}.
Pure SrTiO$_3$ behaves as a band insulator with a gap~\footnote{The superlattice geometry corresponds to the geometry of SrTiO$_3$ and therefore plays no role in this case.} of \SI{\sim 2.4}{\electronvolt}, which is slightly smaller than the experimental value of \SI{\sim 3.2}{\electronvolt}~\cite{Pai_2018}. 
As one layer of LaMnO$_3$ is introduced, and with it $d^4$ centres, $e_g$ states appear within the gap region. The narrow band defined by these states becomes split in occupied and unoccupied regions, leaving Mn in a $t_{2g}^3 e_g^1$ configuration.   
 When more layers of LaMnO$_3$ are added, the $e_g$ states start hybridizing, but their character remain overall insulating. For $n=5$, corresponding to the thickest region of LaMnO$_3$, the $e_g$ band has a width of about \SI{\sim 1.1}{\electronvolt}, while the single sub-bands are even narrower.
 For pure LaMnO$_3$ constrained to the superlattice geometry, a half-metallic FM state is obtained, similarly to what was found for the (001) orientation~\cite{lee_jh-PRB.88.174426,PhysRevB96235112,ZHONG2018406}. The finite conductivity observed in this case highlights the importance of quantum confinement in driving the insulating character, which is not due to strain and symmetry alone.  
 Another interesting feature observed in Figure~\ref{fig:bandos} is the nature of the band gap, which is indirect for $n=1,3,4,5$ and direct for $n=2$. We attribute this dependence to the (111) stacking geometry: in the bilayer limit the Mn sublattice forms a buckled honeycomb lattice, whose symmetry yields a band structure that is highly sensitive to the precise number of layers~\cite{xiao-ncomm2011,weng-PRB.92.195114,tahini-PRB.93.035117,okamoto_jpsj_2018}. 
 Adding or removing one monolayer modifies the symmetry and thus the folding and hybridization of subbands associated with confinement in the growth direction, which can shift the valence- and conduction-band extrema between different $\mathbf{k}$ points. The case for $n=4$ is the most interesting, since the hybridization between two buckled honeycomb bilayers lead to an indirect band gap analogous to the one observed in bilayers of MoS$_2$ and WSe$_2$~\cite{autieri-PhilMag2017,splendiani-NanoL2010.4}. In our superlattices, these effects are partially mitigated by the presence of Jahn-Teller distortions, which modify the perfect buckled honeycomb lattice formed by the bilayer~\cite{jansen2024prm}, thereby reducing the symmetry contrast upon changing $n$. Finally, the band gap decreases when the thickness of LaMnO$_3$ increases, going from \SI{\sim 1}{\electronvolt} for $n=1$ to \SI{\sim 0.5}{\electronvolt} for $n=5$.
%

The layer-resolved van Vleck distortions~\cite{schmitt-PRB.101.214304,vanvleck-JChemPhys1939}, shown in Figure~\ref{fig:angmgn}, are almost entirely of Jahn-Teller type.
Insulating character and Jahn-Teller distortions indicate orbital localisation; rather than forming a metallic band, electrons preferentially localize in a certain layer with a certain orbital type. The analysis of the Mn-$3d$ local density matrices (data not shown) reveals an orbital ordering characterized by linear combinations of $e_g$ orbitals, as depicted in Figure~\ref{fig:bandos}(j). This ordering can also be visualized via the corresponding $e_g$-projected charge densities, shown in Figure~S2 of SM~\cite{supplemental}. Such orbital localization and ordering are hallmarks of correlation-driven and/or Jahn–Teller–assisted physics in Mn-rich systems~\cite{murakami-PRL.81.582,feiner1999prb,okamoto2002prb}. 
The coexistence of these features with FM order makes the present superlattices very different from mixed-valent manganite superlattices where ferromagnetism is commonly discussed as mediated by delocalized carriers~\cite{bhattacharya-PRL.100.257203,nanda-PRB.79.054428,smadici2012prb,myself-npjCM2022,myself-PRB.109.045435}.
Overall, the characteristics exhibited by the present superlattices suggest a localized, strong-coupling regime in which magnetism is governed by an orbital-dependent superexchange \emph{\`a la} Kugel–Khomskii~\cite{kliment-SPU1982,feiner1999prb,Khomskii_2022}.

A more detailed insight can be obtained by calculating the interatomic exchange interactions $J_{ij}$ between the two spins at sites $i$ and $j$. 
To this aim, we employ the full-potential linear muffin-tin orbital (FP-LMTO) method, as implemented in the Relativistic Spin Polarized Toolkit (RSPt)~\cite{RSPt_Springer2010,grechnev07prb76_035107,granas-CMSci2012,kvashnin-PRB.91.125133,lejaeghere-Science2016,kvashnin16prl116}. The calculated exchange interactions are negligible when involving Ti atoms, owing to their near-zero magnetic moments, and also when coupling atoms within the same layer due to the superlattice geometry. Consistent with this, the $J_{ij}$ calculated for $n=1$ are all smaller than 1 meV. For the other compositions, the magnetic coupling is dominated by the nearest-neighbor terms connecting atoms in adjacent (111) layers. Those interactions are plotted along the direction of growth in Figure~\ref{fig:angmgn}, alongside Mn-O-Mn, Ti-O-Mn, Ti-O-Ti bond angles. Each Mn ion is coupled to the trigonal network of the next/previous layer along the Cartesian directions, that is why each plot contains three curves. Moreover, when the GS adopts the $a^-a^-c^+$ tilting pattern, two distinct magnetic sublattices arise~\cite{myself-PRB.109.045435}. 
The reason why the plots are not fully symmetric with respect to a sublattice swap or with respect to the direction of growth is due to having two nonequivalent interfaces (see Figure~\ref{fig:sketch}). 
All superlattices exhibit a strong FM coupling which is mainly due to the $e_g$-$e_g$ components, as shown in Figure~S3 of SM~\cite{supplemental}. In the Kugel--Khomskii picture, the occupied $e_g$ orbitals select the dominant $d$--$p$--$d$ hopping channels and thereby control both the magnitude and sign of the effective exchange~\cite{kliment-SPU1982,feiner1999prb,Khomskii_2022}. 
In our superlattices this manifests most clearly at the interfaces, where the FM coupling in Figure~\ref{fig:angmgn}(c) is the strongest. Deeper in the LaMnO$_3$ region, the FM $e_g$-$e_g$ component decreases and is nearly compensated by the AFM $t_{2g}$--$t_{2g}$ component, which is typically due to an AFM superexchange mechanism in manganites~\cite{dionnebook}. Notice that, for a given composition, the local magnetic moments remain basically constant across all the layers of the superlattice, see Table~S1 of SM~\cite{supplemental}, which implies that the trend of the exchange coupling reflects a change in the $d$--$p$--$d$ hopping channel. This further emphasizes the connection between exchange coupling and cooperative Jahn-Teller distortions in these systems. Looking at the middle panels of Figure~\ref{fig:angmgn}, the bond angles show no evident correlation with the values of the nearest-neighbor exchange interactions. This is consistent with our general conclusion that the exchange coupling originates from a Kugel-Khomskii mechanism, which involves factors beyond the bond angles such as the precise bond length, the orbital texture and Hund's coupling~\cite{kliment-SPU1982,Khomskii_2022}.

In conclusion, we have demonstrated that (111)-oriented superlattices of LaMnO$_3$ and SrTiO$_3$ host an intriguing insulating FM state, characterized by electron localization and orbital ordering. The magnetic order is found to be driven by the orbital-dependent superexchange of Kugel-Khomskii type and is a consequence of the concomitant action of stacking sequence, epitaxial strain and quantum confinement. The band gap exhibits a non-trivial trend with respect to the thickness of the LaMnO$_3$ region, which suggests opportunities for band-gap engineering of insulating ferromagnets for spin-dependent transport.

\vspace{0.5cm}
\begin{acknowledgments}
	\textit{Acknowledgments.} We are thankful to H.-S.\ Kim, U.\ Schwingenschl\"ogl, A.\ Partos, I.\ E.\ Brumboiu, and A.\ Akbari for valuable discussions and technical help. 
    Computational work was performed on resources provided by the National Academic Infrastructure for Supercomputing in Sweden (NAISS), partially funded by the Swedish Research Council through Grant Agreement No. 2022-06725. 
    We additionally appreciate allocation of resources of the Supercomputing Laboratory at King Abdullah University of Science \& Technology (KAUST) in Thuwal, Saudi Arabia. F.\ C. and I.\ D.\ M. acknowledge financial support from the National Research Foundation (NRF) funded by the Ministry of Science of Korea (Grants Nos.\ 2022R1I1A1A01071974 and 2020R1A2C101217411,
	respectively). I.\ D.\ M.\ acknowledges financial support from the European Research Council (ERC), Synergy Grant FASTCORR, Project No.\ 854843. S.\ S. acknowledges funding from the Carl Tryggers Foundation (grant number CTS 22:2013, PI: V.B.). The present project was also supported by the STINT Mobility Grant for Internationalization (Grant No. MG2022-9386). This research is part of the Project No. 2022/45/P/ST3/04247 co-funded by the National Science Center of Poland and the European Union's Horizon 2020 research and innovation program under the Marie Skodowksa-Curie Grant Agreement No. 945339.
\end{acknowledgments}

%

\end{document}


\title{Supplemental material for the manuscript entitled\\ ``Orientation-driven route to an intrinsic insulating ferromagnetic state in manganite superlattices''}

\author{Priyanka Aggarwal}
\thanks{These authors contributed equally to this work.}
\affiliation{Department of Physics and Astronomy, Uppsala University, Box 516, SE-75120, Uppsala, Sweden}
\affiliation{Asia Pacific Center for Theoretical Physics, Pohang, Gyeonbuk 790-784, Republic of Korea}

\author{Kirill B.\ Agapev}
\thanks{These authors contributed equally to this work.}
\affiliation{Physical Science and Engineering Division, King Abdullah University of Science and Technology (KAUST), Thuwal 23955-6900, Kingdom of Saudi Arabia}

\author{Sagar Sarkar}
\affiliation{Department of Physics and Astronomy, Uppsala University, Box 516, SE-75120, Uppsala, Sweden}

\author{Biplab Sanyal}
\affiliation{Department of Physics and Astronomy, Uppsala University, Box 516, SE-75120, Uppsala, Sweden}

\author{Igor Di Marco}
\email[I. Di Marco: ]{igor.dimarco@physics.uu.se}
\affiliation{Institute of Physics, Nicolaus Copernicus University, Grudziadzka 5, 87-100, Toru\'n, Poland}
\affiliation{Department of Physics and Astronomy, Uppsala University, Box 516, SE-75120, Uppsala, Sweden}

\author{Fabrizio Cossu}
\email[F. Cossu: ]{fabrizio.cossu@york.ac.uk}
\affiliation{School of Physics, Engineering and Technology, University of York, Heslington, York YO10 5DD, United Kingdom}
\affiliation{Department of Physics and Institute of Quantum Convergence and Technology, Kangwon National University, Chuncheon, 24341, Republic of Korea}
\affiliation{Department of Physics, School of Natural and Computing Sciences, University of Aberdeen, Aberdeen, AB24 3UE, UK}

\begin{abstract}
In this Supplemental Material, we include a more detailed description of the methodology used for our study, as well as additional data on the charge densities and magnetic moments. Finally, results obtained for a different set of Coulomb interaction parameters are also illustrated.
\end{abstract}


\maketitle 

\newpage
\beginsupplement

\section{Methods and Models}
 Density functional theory (DFT) calculations are performed using the projector-augmented wave method as implemented in the Vienna {\itshape{ab-initio}}
 simulation package (VASP)~\cite{kresse-PRB.54.11169,kresse-PRB.59.1758}; the generalized gradient approximation (GGA) in the Perdew–Burke–Ernzerhof
 parametrisation~\cite{perdew-PRL.77.3865,perdew-PRLerratum.78.1396} is adopted. We used the pseudopotentials for Sr, La, Mn, Ti and O which treat explicitly
 10, 11, 15, 12 and 6 electrons, respectively and we chose an energy cutoff on the plane-wave of \SI{550}{\electronvolt}. Systems investigated include (111)-oriented superlattices of chemical formula  (SrTiO$_3$)$_{6–n}\vert$(LaMnO$_3$)$_{n}$ and $n$ ranging from $0$ to $6$. The two end points correspond to the bulk materials SrTiO$_3$ and LaMnO$_3$ constrained to the geometry of the superlattice.  
 The overall thickness of 6 layers represents the minimal repeat unit compatible with the relevant rhombohedral and orthorhombic symmetries, which usually characterize LaMnO$_3$ in superlattices~\cite{myself-npjCM2022}.
 The sampling of the Brillouin Zone is performed with a $\Gamma$-centred $\mathbf{k}$-mesh of $7\times 4\times 3$ points for all superlattices. A Gaussian smearing of \SI{10}{\milli\electronvolt} is used for all calculations except for those to obtain the density of states (DOS), for which the tetrahedron method is adopted. An energy tolerance of \SI{1e-6}{\electronvolt} is chosen for the convergence of the  electron density loop during the structural optimisation as well as the calculations of the electronic properties. Ionic positions are considered relaxed when forces are smaller than \SI{5e-3}{\electronvolt\per\angstrom}. 
 Starting from the relaxed lattice constant of cubic SrTiO$_3$, namely $a_{STO}=$~\SI{3.925}{\angstrom}, (111)-oriented supercells with six transition-metal layers were constructed, yielding orthogonal lattice vectors of $\sqrt{2} a_{STO}$,  $\sqrt{6} a_{STO}$ and  $2 \times \sqrt{3} a_{STO}$. The in-plane vectors were kept fixed to simulate coherent epitaxy, while the out-of-plane lattice parameter was optimized for each composition.

 Due to the well-known problems of naive DFT in describing localized $3d$ states in magnetic oxides~\cite{kotliar2006,varignon-ncomm2019}, these shells in Mn and Ti are treated within the rotationally invariant DFT+$U$ approach by Liechtenstein \textit{et al.}~\cite{liechtenstein-PRB.52.r5467}. The Coulomb interaction parameters for the Mn-$3d$ states are set to $U =$ \SI{4.0}{\electronvolt} and $J =$ \SI{1.0}{\electronvolt}, which were previously used to investigate analogous superlattices~\cite{nanda-PRB.81.224408,nanda-PRL.101.127201,chen_zh-PRL.119.156801,myself-EPL2017,myself-npjCM2022,jansen2024prm,liu2025prb} and are within the range reproducing the bulk properties to a good accuracy~\cite{mellan-PRB.92.085151,Banerjee2019prb,myself-npjCM2022}. For completeness, we also perform another set of calculations with $U =$ \SI{5.0}{\electronvolt} and $J =$ \SI{1.0}{\electronvolt}, whose results are shown below. The Coulomb interaction parameters for the Ti-$3d$ states are set to $U =$ \SI{5.0}{\electronvolt} and $J =$ \SI{0.5}{\electronvolt}, which is also in line with previous literature~\cite{PhysRevLett.116.157203,PhysRevB.96.161409,PhysRevLett.119.256403}. The double counting term is described within the fully localized limit (FLL)~\cite{kotliar2006,liechtenstein-PRB.52.r5467}. These settings were also used to obtain the relaxed lattice constant $a_{STO}$.

 The superlattices are built from the $Pnma$ and $R\bar{3}c$ bulk structures, featuring $a^{-}a^{-}c^{+}$ and $a^{-}a^{-}a^{-}$ tilting systems, respectively. Describing the former tilting system requires a doubling of the in-plane periodicity with respect to the latter
 one, for (111)-oriented superlattices; this corresponds to 2 formula units per layer. For an accurate comparison of energy with the same spacing of reciprocal lattice points, we nevertheless model the structure with $a^{-}a^{-}a^{-}$ tilting systems in the same $Pnma$ supercell. This means that all the results reported here required the calculations of supercells with 60 atoms. 
 The superlattices are modeled with planes stacked along the (111) direction and are oriented as in our previous study on superlattices of LaMnO$_3$ and SrMnO$_3$ \cite{myself-PRB.109.045435}, where the octahedral axes $(a, b, c)$ can be expressed in terms of the Cartesian coordinates $(x, y, z)$ as $(\frac{1}{\sqrt{2}} x + (\sqrt{2} - 1) y + \frac{2\sqrt{3} - \sqrt{6}}{2} z;
 \frac{-1}{\sqrt{2}} x + (\sqrt{2} - 1) y + \frac{2\sqrt{3} - \sqrt{6}}{2} z;
 (2 - 2\sqrt{2})y + (\sqrt{6} - \sqrt{3}) z)$. Finally, the images of the structures are produced with VESTA JP-Minerals \cite{VESTA-JACr2011}, and the analysis of the electronic properties is performed with the aid of the post-processing code VASPKIT \cite{VASPKIT-1908.08269}.

The structural distortions are calculated using the same notation proposed by van Vleck~\cite{vanvleck-JChemPhys1939}. The volume-breathing distortions $Q_1$ and the volume-conserving Jahn-Tellers distortions $Q_2$ and $Q_3$ are defined according to the following equations~\cite{schmitt-PRB.101.214304,myself-npjCM2022}:
\begin{align}
\label{eq:mode1}
    Q_1 &= \frac{1}{\sqrt{3}} \left( \Delta x + \Delta y + \Delta z \right)\\
\label{eq:mode2}
    Q_2 &= \frac{1}{\sqrt{2}} \left( \Delta x - \Delta y \right)\\
\label{eq:mode3}
    Q_3 &= \frac{1}{\sqrt{6}} \left(-\Delta x - \Delta y + 2\Delta z \right).
\end{align}
where the variations refer to the octahedral lengths $x$, $y$, $z$ with respect to their average values. Because the modes propagate along the (111) crystallographic direction, the full notation for equations (\ref{eq:mode1}-\ref{eq:mode3}) should be $Q^R_1$, $Q^R_2$ and $Q^R_3$. However, since there is no other direction of propagation for the modes, a simpler notation is used throughout the manuscript. Layered-resolved charge and spin distributions are instead computed according to Bader theory~\cite{bader-AccChemRes1985}.

Following the structural and magnetic analysis made with VASP, further calculations for the ground-states are performed by means of the full-potential linear muffin-tin orbital (FP-LMTO) method, as implemented in the Relativistic Spin Polarized Toolkit (RSPt)~\cite{RSPt_Springer2010}. This approach provides more accuracy, although at a higher computational cost, since it is an all-electron theory~\cite{lejaeghere-Science2016}.
In our calculations,  the basis functions include two distinct energy sets with three different tail energies, namely 0.3 Ry, -2.3 Ry, and -1.5 Ry. All three tail energies are employed for the $s$ and $p$ states, while only the first two tail energies are applied to the other states included in the basis set. Valence states comprehend $6s$, $6p$, $5d$, and $4f$ for La; $5s$, $5p$, $4d$, and $4f$ for Sr; $4s$, $4p$, $3d$ for Mn and Ti; and $2s$, $2p$ for O. Semicore states include $5s$, $5p$, $5d$ and $4f$ for La; $4s$, $4p$, $4d$ and $4f$ for Sr; and $3s$, $3p$, $3d$ for Mn and Ti. The muffin-tin radii are chosen as 2.50 a.u. for La and Sr, 1.80 a.u. for Mn and Ti and 1.70 a.u. for O. The Brillouin zone sampling is performed with the same $\mathbf{k}$-mesh as the one used for VASP to ensure the best compatibility. In line with the VASP setup, the DFT+U correction was applied to the Mn-$3d$ and Ti-$3d$ states only, using the same Coulomb interaction parameters and with the double counting treated in the FLL. The local orbital basis used for the DFT+U correction as well as for the analysis of the interatomic exchange is constructed from the so-called muffin-tin heads, as described in previous works~\cite{grechnev07prb76_035107,granas-CMSci2012,kvashnin-PRB.91.125133}
The electronic structure calculations with RSPt are based on the VASP relaxed structures and are intended to provide an accurate analysis of the magnetic coupling. The layer-resolved magnetic moments and projected DOS obtained with the two codes are in very good agreement, which ensured the consistency of our results. 
The inter-atomic exchange parameters $J_{ij}$ between the two spins at sites $i$ and $j$ are then calculated by mapping the magnetic excitations onto an effective Heisenberg model whose Hamiltonian reads as
 \begin{equation}
     \hat{H}=-\sum_{i\neq j} J_{ij} \overrightarrow{e_i} . \overrightarrow{e_j}
 \end{equation}
 where $\overrightarrow{e_i}$ and $\overrightarrow{e_j}$ are unit vectors along the direction of the magnetic moments at the corresponding sites. In RSPt, the $J_{ij}$'s are calculated by employing a generalized version of the magnetic force theorem~\cite{kvashnin-PRB.91.125133}. The orbital-decomposition is performed through the implementation of Ref.~\onlinecite{kvashnin16prl116}.

\newpage
\section{Competing magnetic orders}
\begin{figure*}[b!]
\vspace{0.4cm}
\centering
    \includegraphics[width=0.95\linewidth]{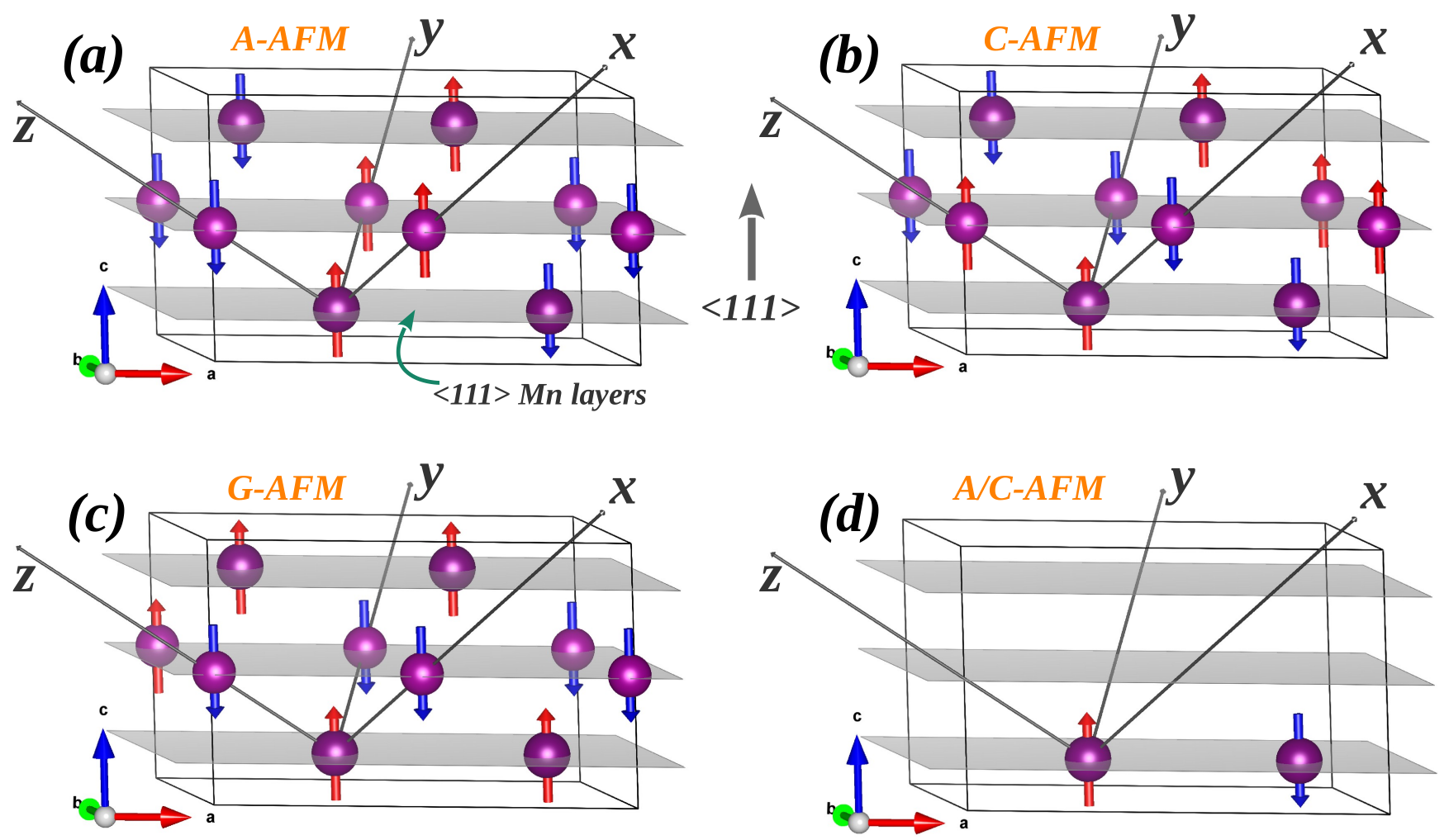}
	\caption{Schematic depiction of AFM orders of A-type (a),  C-type (b), and G-type (c) in the (111)-oriented superlattice geometry of the Mn sublattice. The Mn atoms are represented as purple spheres arranged on the (111) planes (gray shaded areas). The three orthogonal lines represent the pseudocubic $x$, $y$, and $z$ directions, corresponding to the nearest-neighbor Mn-Mn bonds. The red and blue arrows indicate up- and down-spins, respectively. A more detailed view of these magnetic orders is also provided in our previous article~\cite{myself-PRB.109.045435}. In the limit of only one-layer of Mn atoms (d), only one unique 'up-down' AFM configuration is possible.}
\label{fig:magpatterns}
\end{figure*}
We illustrate how the conventional collinear anti-ferromagnetic (AFM) orders of A-, C-, and G-type manifest within the (111)-oriented superlattice geometry of the Mn sublattice. 
These three AFM orders, which are labeled as A-AFM, C-AFM and G-AFM for simplicity, are shown in Figure~\ref{fig:magpatterns}, in panels (a), (b), and (c), respectively. In these images, the Mn atoms are represented as purple spheres arranged on the (111) planes (gray shaded areas). The three orthogonal lines represent the pseudocubic $x$, $y$, and $z$ directions, corresponding to the nearest-neighbor Mn-Mn bonds. The red and blue arrows indicate up- and down-spins, respectively. A more detailed view of these magnetic structures is also provided in our previous article~\cite{myself-PRB.109.045435}.
Figure~\ref{fig:magpatterns}(d) highlights the unique case of a single Mn layer ($n=1$) along the (111) direction. In this two-dimensional limit, the out-of-plane pseudocubic bond connectivity is broken. Consequently, the distinction between A-, C-, and G-type orders effectively vanishes. In this single-layer limit, only one unique 'up-down' AFM configuration is possible, which explains the degeneracy reported on Table~I of the main article.  Therefore, in our study, we refer to this state as a generic AFM state, interchangeably as A- or C-type.

\section{Orbital ordering}
\begin{figure*}[b!]
\centering
    \includegraphics[width=0.95\linewidth]{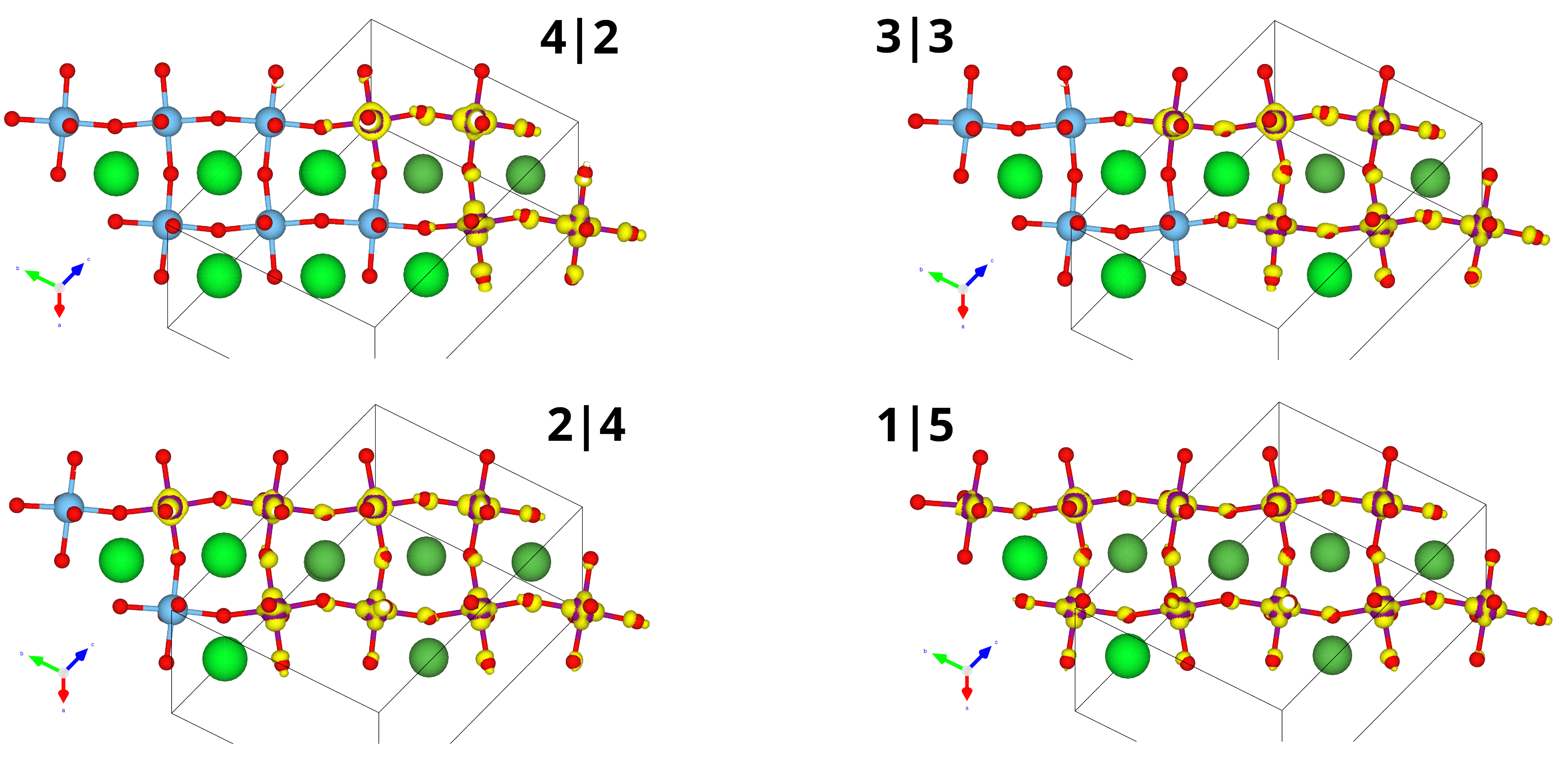}
	\caption{Charge density projected over the $e_g$ orbitals for (111)-oriented (SrTiO$_3$)$_{6–n}\vert$(LaMnO$_3$)$_{n}$ superlattices for $n=2,3,4,5$. For each superlattice, the ferromagnetic (FM) ground state (GS) in the $a^-a^-c^+$ tilting pattern is considered, as described in the main article.}
\label{fig:chargedens}
\end{figure*}
The charge density depicting the $e_g$ states in the FM GS of (111)-oriented (SrTiO$_3$)$_{6–n}\vert$(LaMnO$_3$)$_{n}$ superlattices for $n=2,3,4,5$ is shown in Figure~\ref{fig:chargedens}. Orbital ordering of the $e_g$ states is clearly visible across the layers, for all considered thicknesses. 

\newpage
\section{Magnetic moments}
\begin{table}[b!]
    \centering
    \begin{tabular}{ccccccc}
      \:\:\:\: SL \:\:\:\: & \:\:\:\: Layer 1 \:\:\:\: & \:\:\:\: Layer 2 \:\:\:\: & \:\:\:\: Layer 3 \:\:\:\: & \:\:\:\: Layer 4 \:\:\:\: & \:\:\:\: Layer 5 \:\:\:\: & \:\:\:\: Layer 6 \:\:\:\:\\
       \hline
       5-1  & 0.05 \: 0.05 & 0.00 \: 0.00 & 0.00 \: 0.00 & 0.00 \: 0.00 & 0.04 \: 0.03 & 3.52 \: 3.51 \\
       4-2  & 0.04 \: 0.04 & 0.00 \: 0.00 & 0.00 \: 0.00 & 0.05 \: 0.05 & 3.64 \: 3.62 & 3.72 \: 3.75 \\
       3-3  & 0.04 \: 0.04 & 0.00 \: 0.00 & 0.05 \: 0.05 & 3.69 \: 3.68 & 3.68 \:3.68 & 3.71 \: 3.72 \\
       2-4  & 0.05 \: 0.05 & 0.06 \: 0.06 & 3.70 \: 3.70 & 3.70 \: 3.70 & 3.70 \: 3.69 & 3.75 \: 3.77 \\
       1-5  & 0.09 \: 0.09 & 3.77 \: 3.74 & 3.68 \: 3.72 & 3.71 \: 3.70 & 3.72 \: 3.69 & 3.72 \: 3.76\\ \\
    \end{tabular}
    \caption{Spin magnetic moments projected at the transition metal sites for  (111)-oriented superlattices of chemical formula (SrTiO$_3$)$_{6-n}\vert$(LaMnO$_3$)$_{n}$, where $n=1-5$. For $n=2,3,4,5$, the FM ground state is considered. For $n=1$, the FM state is considered as well, for sake of comparison. See also the discussion in the main text. For each superlattice (SL), two values (in $\mu_B$) are reported per each layer, corresponding to two different sublattices that develop for the $a^{-}a^{-}c^{+}$ tilting pattern~\cite{myself-PRB.109.045435}. The sublattices for $n=1$ are degenerate because this periodicity prefers the $a^{-}a^{-}a^{-}$ tilting pattern. Two different interfaces are present in each system, namely Ti{$\vert$}SrO$_3${$\vert$}Mn{$\vert$}LaO$_3$ and Mn{$\vert$}LaO$_3${$\vert$}Ti{$\vert$}SrO$_3$. The latter is shown always between layers 6 and 1, while the former scales accordingly to the superlattice composition.}
    \label{tab:magmom}
\end{table}
The values of the projected magnetic moments at the transition metal sites are reported in the Table~\ref{tab:magmom} for all the investigated superlattices in their FM GS, as defined in the caption as well as in the main text. Two different interfaces are present in each system, namely Ti{$\vert$}SrO$_3${$\vert$}Mn{$\vert$}LaO$_3$ and Mn{$\vert$}LaO$_3${$\vert$}Ti{$\vert$}SrO$_3$. The latter is shown always between layers 6 and 1, while the former scales accordingly to the superlattice composition. Notice that for each thickness, there is no significant variation of the size of the magnetic moments across the layers. This is a further indication of orbital localization, as visible in Figure~\ref{fig:chargedens} and discussed in the main manuscript.

\section{Orbital decomposition of the exchange coupling}
\begin{figure*}[b!]
\centering
	\includegraphics[trim = 0 0 0 0,width=1.0\linewidth]{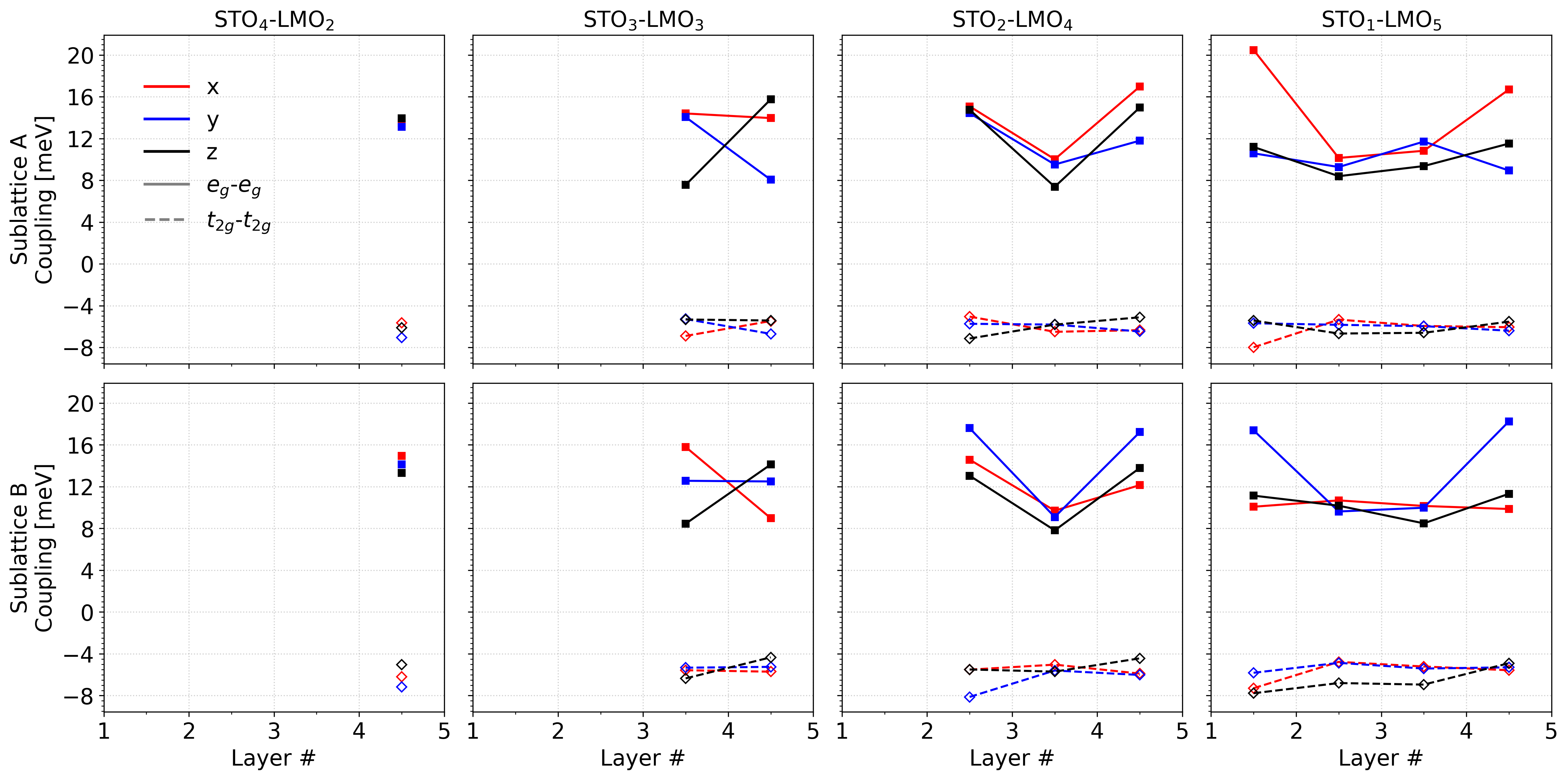}
	\caption{Orbital-decomposed nearest-neighbor magnetic interactions in (111)-oriented (SrTiO$_3$)$_{6–n}\vert$(LaMnO$_3$)$_{n}$ superlattices for $n=2,3,4,5$. For each superlattice, the FM GS in the $a^-a^-c^+$ tilting pattern is considered, as described in the main article. Two sublattices are shown, since they emerge from the $a^-a^-c^+$ tilting pattern }
\label{fig:orbmag}
\end{figure*}
The values of the orbital-decomposed nearest-neighbor magnetic interactions in (111)-oriented (SrTiO$_3$)$_{6–n}\vert$(LaMnO$_3$)$_{n}$ superlattices for $n=2,3,4,5$ are shown in Figure~\ref{fig:orbmag}. These data refer to the FM GS, which adopts the $a^-a^-c^+$ tilting pattern, and thus two sublattices emerge~\cite{myself-PRB.109.045435}. The AFM coupling due to the $t_{2g}$-$t_{2g}$ channel shows a less varying behavior than the FM coupling due to the $e_{g}$-$e_{g}$ channel. The mixed $e_{g}$-$t_{2g}$ components have negligible values and are therefore not plotted. As for the magnetic moments in Table~\ref{tab:magmom}, two different interfaces are present in each system, namely Ti{$\vert$}SrO$_3${$\vert$}Mn{$\vert$}LaO$_3$ and Mn{$\vert$}LaO$_3${$\vert$}Ti{$\vert$}SrO$_3$. The latter is shown always between layers 6 and 1, while the former scales accordingly to the superlattice composition.

\section{Dependence on the Coulomb interaction parameters}
\begin{table}[b!]
    \centering
    \begin{tabular}{cccccc}
 \:\:\:\:\:\: $a^-a^-a^-$  \:\:\:\:\:\: &  \:\:\:\:\:\: 5-1  \:\:\:\:\:\: &  \:\:\:\:\:\: 4-2  \:\:\:\:\:\: &  \:\:\:\:\:\: 3-3  \:\:\:\:\:\: &  \:\:\:\:\:\: 2-4  \:\:\:\:\:\: &  \:\:\:\:\:\: 1-5 \:\:\:\:\:\: \\
\hline
 A-AFM &    GS    &   6.29  &  13.00  &  17.55  &  41.02  \\
 C-AFM &    GS    &   5.81  &  12.10  &  28.84  &  36.65  \\
    FM &   0.58   &    GS   &   3.41  &  11.87  &  18.72  \\
 G-AFM &    ---   &  12.76  &  29.10  &  42.45  &  62.61  \\ \\
 \:\:\:\: $a^-a^-c^+$ \:\:\:\:\:\: &  \:\:\:\:\:\: 5-1  \:\:\:\:\:\: &  \:\:\:\:\:\: 4-2  \:\:\:\:\:\: &  \:\:\:\:\:\: 3-3  \:\:\:\:\:\: &  \:\:\:\:\:\: 2-4  \:\:\:\:\:\: &  \:\:\:\:\:\: 1-5 \:\:\:\:\:\: \\
\hline
 A-AFM &  10.45  &   4.32  &   4.99  &   6.42  &   9.57  \\
 C-AFM &  10.45  &   6.83  &  23.56  &  20.22  &  37.25  \\
    FM &  10.96  &   0.96  &    GS   &    GS   &    GS    \\
 G-AFM &  10.96  &  15.90  &  26.91  &  34.98  &  44.79  \\ \\
    \end{tabular}
    \caption{Relative energy of competing magnetic states in the most favorable tilting patterns for (111)-oriented superlattices of chemical formula (SrTiO$_3$)$_{6-n}\vert$(LaMnO$_3$)$_{n}$, where $n =1-5$, as calculated by using $U = 5.0$~eV and $J = 1.0$~eV as Coulomb interaction parameters form the Mn-$3d$ states. Values are given in \SI{}{meV} per formula unit and with respect to the GS for any given $n$. The $a^{-}a^{-}c^{+}$ ($a^{-}a^{-}a^{-}$) tilting pattern corresponds to the $Pnma$ ($R\bar{3}c$) crystal structure.}
    \label{tab:totenergies}
\end{table}
In Table~\ref{tab:totenergies}, the relative energies of competing magnetic states in (111)-oriented superlattices (SrTiO$_3$)$_{6–n}\vert$(LaMnO$_3$)$_{n}$ with $n =1-5$ are shown, as calculated by using $U = 5.0$~eV and $J = 1.0$~eV for the Mn-$3d$ states. The rest of the computational parameters are identical to those used in the main manuscript. Notice that for $n=1$ the A-AFM and C-AFM orders become equivalent, as shown in Figure~\ref{fig:magpatterns}(d). These values should be compared with those reported in Table~I of the main manuscript. A comparison between those two datasets demonstrates that a small variation of the Coulomb interaction parameters that is compatible with available literature~\cite{PhysRevLett.116.157203,PhysRevB.96.161409,PhysRevLett.119.256403} does not change the physical picture substantially. As expected from the detailed analysis by Mellan \textit{et al.}~\cite{mellan-PRB.92.085151}, this larger value of the Hubbard term tends to favor FM interactions, leading to a FM GS in bulk LaMnO$_3$ as well. As a result, in Table~\ref{tab:totenergies}, the difference between the energies of the AFM GS and FM first excited state for the superlattice with $n=1$ becomes almost negligible. A further increase of $U$ is expected to drive a magnetic phase transition to FM order even for the $n=1$ case. Another noticeable difference with our default settings is that in Table~\ref{tab:totenergies}, the $a^{-}a^{-}a^{-}$ tilting pattern becomes the GS for $n=2$ as well, although the system still remains in a FM insulating state. A final important aspect is that this larger $U$ applied to the Mn-$3d$ states helps stabilizing the solutions for $n=1$ in  the $a^{-}a^{-}c^{+}$ tilting pattern. We believe that this is driven by the higher stability of the high-spin state, due to a stronger Coulomb repulsion, as emphasized by Mellan \textit{et al.}~\cite{mellan-PRB.92.085151}.
Thus, these structures do not collapse into the $a^{-}a^{-}a^{-}$ tilting pattern as it happens for a weaker Coulomb repulsion term. Finally, all the GS reported in Table~\ref{tab:totenergies} have an insulating character. Overall, the data presented in this section and in the main manuscript demonstrate the robustness of our prediction of a FM insulating GS for most compositional parameters.

%
